\documentclass[journal,twoside,web]{ieeecolor}
\usepackage{lcsys}
\usepackage{textcomp}
\usepackage{cite}

\usepackage{amsthm}
\usepackage{amsmath,amssymb,amsfonts,bm,mathrsfs}
\usepackage{graphicx}
\usepackage[hidelinks]{hyperref}
\usepackage{cleveref}
\usepackage{tikz}
\usepackage{algorithm}
\usepackage{algpseudocode}
\usepackage[dvipsnames]{xcolor}
\usetikzlibrary{arrows.meta,positioning,calc,intersections,fit,backgrounds}
\usepackage[caption=false,font=footnotesize]{subfig}

\usepackage{microtype}
\usepackage{orcidlink}

\allowdisplaybreaks
\newtheorem{lemma}{Lemma}

\newtheorem{theorem}{Theorem}

\newtheorem{assumption}{Assumption}

\title{Learning-Based Shrinking Disturbance-Invariant Tubes for State- and Input-Dependent Uncertainty}

\author{Abdelrahman Ramadan\textsuperscript{\large \orcidlink{0000-0001-5396-4768}}, \IEEEmembership{Graduate Student Member, IEEE}\textsuperscript{1}, Sidney Givigi\textsuperscript{\large \orcidlink{0000-0002-3829-3545}}, \IEEEmembership{Senior Member, IEEE}\textsuperscript{2}
        \thanks{\textsuperscript{1}A. Ramadan is with Electrical and Computer Engineering (ECE), Smith Engineering and with Ingenuity Labs Research Institute, Queen’s University, Kingston, ON K7L 3N6 Canada, 20amr3@queensu.ca}
        \thanks{\textsuperscript{2}S. Givigi is with the School of Computing and with Ingenuity Labs Research Institute, Queen’s University, Kingston, ON K7L 3N6 Canada, sidney.givigi@queensu.ca }}

\begin{document}

\maketitle

\thispagestyle{empty}  
\pagestyle{empty}       


\begin{abstract}
We develop a learning-based framework for constructing shrinking disturbance-invariant tubes under state- and input-dependent uncertainty, intended as a building block for tube Model Predictive Control (MPC), and certify safety via a lifted, \textit{isotone} (order-preserving) fixed-point map. Gaussian Process (GP) posteriors become $(1-\alpha)$ credible ellipsoids, then polytopic outer sets for deterministic set operations. A two-time-scale scheme separates \emph{learning epochs}, where these polytopes are frozen, from an inner, outside-in iteration that converges to a compact fixed point $Z^\star\!\subseteq\!\mathcal G$; its state projection is RPI for the plant. As data accumulate, disturbance polytopes tighten, and the associated tubes nest monotonically, resolving the circular dependence between the set to be verified and the disturbance model while preserving hard constraints. A double-integrator study illustrates shrinking tube cross-sections in data-rich regions while maintaining invariance.

\end{abstract}

\begin{keywords}
Data-driven control, set invariance, MPC.
\end{keywords}


\section{Introduction}

Robust Model Predictive Control (MPC) fundamentally relies on set-theoretic invariance to guaranty safety. Classical results provide disturbance-invariant and Robust Positively Invariant (RPI) sets, as well as practical computation for \emph{fixed}, state-independent bounds~\cite{blanchini2008set,kolmanovsky1998theory}. While these constructions are mature, they rely on worst-case bounds, and are therefore conservative. To reduce this conservatism, tube MPC has evolved to \emph{state-dependent} cross-sections that adapt to local state disturbance characteristics~\cite{ghaemi2011robust,schaich2015robust}, to formulations that capture state–input dependent effects in practice~\cite{8815343}, and to parametric RPI sets that represent input-scaled uncertainty~\cite{darup2016parametric}. Recent work improves computational tooling for state-/input-dependent invariance~\cite{https://doi.org/10.1002/rnc.5688}.

In parallel, learning-based MPC approaches have been gaining a lot of traction, particularly those based on Gaussian Processes (GPs) that model unknown dynamics and infer residual disturbances from data. GPs provide nonparametric posteriors over model mismatch~\cite{rasmussen2006gaussian}; GP–MPC has leveraged these posteriors for safe exploration~\cite{koller2018learning}, to translate confidence sets into constraint tightenings~\cite{hewing2020cautious}, and to develop numerically tractable controllers~\cite{10886350}. Extensions consider distributionally robust treatments~\cite{10753052} and adaptive/online updates~\cite{10604720}. However, most GP–MPC approaches reason via chance constraints or tightenings: they do not deliver an invariant-set synthesis that \textbf{(i)} treats joint $(x,u)$-dependent, \emph{learned} disturbance sets and \textbf{(ii)} yields tubes that nest monotonically across learning epochs with explicit control of representation complexity. Relatedly, while reachability for LTI systems with moving (state- and input-dependent) disturbance sets is well recognized~\cite{rakovic2003reachability}, available relaxations (e.g., state-/input-dependent tubes and tools~\cite{ghaemi2011robust, https://doi.org/10.1002/rnc.5688}) do not offer a convergent invariant-set iteration beyond low dimension---precisely the gap we target. \textit{Motivation for state and input dependence.} Many ubiquitous effects scale with \emph{both} state and input: aerodynamic drag and lift vary with velocity (and angle), actuator efficiency and rate limits scale with the commanded input, and friction/contact forces depend on pose and normal load. A single worst-case bound is thus overly conservative; locally sized, state-/input-dependent sets tighten where data are informative.
Following the work in~\cite{rakovic2003reachability}, other works show continued relevance through state-dependent tubes~\cite{schaich2015robust}, input-scaled/parametric RPI sets~\cite{darup2016parametric}. We follow this line but replace hard-specified dependence with \emph{learned} $(x,u)$-dependent bounds.

In line with GP–MPC practice, we adopt an LTI nominal model and learn the residual with a GP~\cite{koller2018learning,hewing2020cautious}.  
However, extending the state- and input-dependent invariant-set computation to fully non-linear plants would require non-convex reachable-set propagation or differential-inclusion machinery with stronger regularity and substantially higher computational load. We leave this for future work  (LPV/local-linear variants fit naturally into our pipeline).

\textit{Novelty in context.}
Relative to state- and input-dependent tubes, invariance tools~\cite{ghaemi2011robust,schaich2015robust,8815343,darup2016parametric} and GP–MPC tightenings~\cite{koller2018learning,hewing2020cautious,10886350,9992469,wang2024tube,10604720}, we: \textbf{(i)} wrap GP posteriors into polytopic $(1-\alpha)$ confidence sets that are frozen per epoch to avoid circularity; \textbf{(ii)} lift the plant into a fixed-graph space and show a monotone outside–in iteration whose state projection yields RPI tubes; and \textbf{(iii)} treat scalability via support-function approximations and anchor grids that bound facet growth, ensuring tubes nest as data contract the learned bounds. This complements GP–MPC: we certify safety by invariant sets under learned $(x,u)$-dependent uncertainty rather than solely by chance-constrained tightenings.

This letter is organized as follows: Section \ref{sec:RPI_compute} develops our approach for RPI sets computation under learned state- and input-dependent disturbances. Section \ref{sec:Synthesis} presents our control synthesis. Sections \ref{sec:simulations} and \ref{sec:conclusion} present simulation results and conclusions respectively.

\section{RPI Computation under Learned State- and Input-Dependent Disturbances} \label{sec:RPI_compute}

The computation of RPI sets for LTI systems with state- and input-dependent uncertain unmodeled dynamics (henceforth referred to as ``\textit{disturbances}'') represents a fundamental challenge in safe autonomy as traditional robust control methods assume fixed, state-independent disturbance sets.
In this section, we consider RPI sets for disturbances that depend on the system state and input.

\subsection{State- and Input-Dependent Disturbances}\label{subsec:2A}

Let us consider a general discrete-time LTI system with state- and input-dependent disturbances:
\begin{equation}
    \mathbf{x}(k+1) = \mathbf{A}\mathbf{x}(k) + \mathbf{B}\mathbf{u}(k) + \mathbf{w}(\mathbf{x}(k),\mathbf{u}(k)),
\label{eq:system}
\end{equation}
\noindent where $\mathbf{x}(k) \in \mathbb{X} \subseteq \mathbb{R}^n$ is the system state, $\mathbf{u}(k) \in \mathbb{U} \subseteq \mathbb{R}^m$ is the control input, with $\mathbb{X}$ and $\mathbb{U}$ being compact convex constraint sets, $\mathbf{A} \in \mathbb{R}^{n \times n}$ and $\mathbf{B} \in \mathbb{R}^{n \times m}$ are the known system matrices, and $\mathbf{w}(\mathbf{x},\mathbf{u}) \in \mathbb{R}^n$ represents the state- and input-dependent disturbance \cite{rakovic2003reachability}.

The component $\mathbf{w}(\mathbf{x},\mathbf{u})$ introduces coupling between the state trajectory and the command input in a stochastic fashion. Since the true disturbance function $\mathbf{w}(\mathbf{x},\mathbf{u})$ is unknown, we employ a data-driven approach to learn this mapping from observed system behavior. The disturbance uncertainty set will be specified in Section~\ref{subsec:2B} from the GP posterior (mean and covariance) at $(\mathbf{x},\mathbf{u})$, where $\hat{\boldsymbol{\mu}}_\mathbf{w}(\mathbf{x},\mathbf{u})$ and $\hat{\boldsymbol{\Sigma}}_\mathbf{w}(\mathbf{x},\mathbf{u})$ are data-driven estimates of the disturbance mean and covariance at state-input pair $(\mathbf{x},\mathbf{u})$, and $\alpha$ controls the confidence level of the uncertainty set. 

\subsection{Learning-Based Disturbance Modeling}\label{subsec:2B}

We employ GPs to learn the unknown state- and input-dependent disturbance mapping $\mathbf{w}(\mathbf{x},\mathbf{u})$ from observed system trajectories. GPs provide a framework for uncertainty quantification, yielding posterior distributions that naturally capture both aleatoric uncertainty (measurement noise) and epistemic uncertainty (model uncertainty). This enables adaptive confidence bounds that shrink in data-rich regions while maintaining conservative estimates in unexplored areas, a property essential for our shrinking tube MPC framework.
We begin by collecting system trajectory data to construct the disturbance dataset.

\textbf{Data Collection:} Using system trajectories, we compute the disturbance at each time step as the model mismatch:
\begin{equation}
\mathbf{w}^{(j)} = \mathbf{x}^{(j+1)} - \mathbf{A}\mathbf{x}^{(j)} - \mathbf{B}\mathbf{u}^{(j)}, \quad j = 1, \ldots, N_{\mathrm{data}}.
\label{eq:w}
\end{equation}
If we define $\mathbf{z}=(\mathbf{x},\mathbf{u})$, this yields a dataset $\mathcal{D} = \{(\mathbf{z}^{(j)}, \mathbf{w}^{(j)})\}_{j=1}^{N_{\mathrm{data}}}$ of state-input-disturbance triplets.

\textbf{GP Model Structure:} We model each component of the disturbance vector independently to maintain computational tractability:
\begin{equation*}
w_i(\mathbf{x},\mathbf{u}) \sim \mathcal{GP}\bigl(0, k_i((\mathbf{x},\mathbf{u}), (\mathbf{x}',\mathbf{u}'))\bigr), \quad i = 1, \ldots, n
\end{equation*}
where $\mathcal{GP}(\mu(\cdot), k(\cdot,\cdot))$ denotes a Gaussian Process with mean function $\mu(\cdot)$ and covariance function $k(\cdot,\cdot)$, with $k_i(\cdot,\cdot)$ as the covariance kernel for the $i$-th component.

\begin{assumption}[Component Independence]
    The disturbance components are conditionally independent given the state-input pair, i.e., $\text{Cov}[\mathbf{w}_i(\mathbf{x},\mathbf{u}), \mathbf{w}_j(\mathbf{x},\mathbf{u})] = 0$ for $i \neq j$. 
\end{assumption} 
This assumption simplifies computation while often providing reasonable approximations for many physical systems where coupling between disturbance components is weak.

\textbf{Kernel Function:} For each disturbance component \(w_i\) we use the product kernel “RBF\(\times\)ExpSineSquared” with white noise, written concisely as
\(k_i\big((\mathbf{x},\mathbf{u},t),(\mathbf{x}',\mathbf{u}',t')\big)=\sigma_{f,i}^2\,k_{\mathrm{SE}}(\mathbf{z},\mathbf{z}';\ell_i)\,k_{\mathrm{PER}}(t,t';p_i,\ell_{p,i})+\sigma_{n,i}^2\,\delta_{(\mathbf{x},\mathbf{u},t),(\mathbf{x}',\mathbf{u}',t')}\),
where \(\mathbf{z}=[\mathbf{x}^\top,\mathbf{u}^\top]^\top\), \(k_{\mathrm{SE}}(\mathbf{z},\mathbf{z}';\ell_i)=\exp(-\|\mathbf{z}-\mathbf{z}'\|^2/(2\ell_i^2))\) is the squared–exponential (RBF) over \((\mathbf{x},\mathbf{u})\), \(k_{\mathrm{PER}}(t,t';p_i,\ell_{p,i})=\exp\!\big(-2\sin^2(\pi (t-t')/p_i)/\ell_{p,i}^2\big)\) is the ExpSineSquared factor on the scalar periodic feature \(t\), \(\sigma_{f,i}^2\) is the signal variance, \(\ell_i\) the isotropic length–scale on \((\mathbf{x},\mathbf{u})\), \(p_i\) the period, \(\ell_{p,i}\) its smoothness scale, \(\sigma_{n,i}^2\) the i.i.d.\ white–noise variance, and \(\delta_{\cdot,\cdot}\) the Kronecker delta.

\textbf{From Data to GP Model:} Given observed trajectories, we compute disturbances as in~\eqref{eq:w},
yielding the dataset $\mathcal{D} = \{(\mathbf{x}^{(j)}, \mathbf{u}^{(j)}, \mathbf{w}^{(j)})\}_{j=1}^{N_{\mathrm{data}}}$. 
We model each disturbance component independently as $w_i(\mathbf{x},\mathbf{u}) \sim \mathcal{GP}(0, k_i)$, resulting in posterior mean 
$\hat{\boldsymbol{\mu}}_\mathbf{w}(\mathbf{x},\mathbf{u}):\mathbb{X}\times\mathbb{U}\rightarrow\mathbb{R}^n$ where each component $i$ is $\hat{\mu}_{w,i}(\mathbf{z})= \mathbf{k}_i(\mathbf{z},Z)^\top\bigl(\mathbf{K}_i+\sigma^2_{n,i}\mathbf{I}\bigr)^{-1}\mathbf{w}_i,$ where
$Z = \{\mathbf{z}^{(j)}\}_{j=1}^{N_{\mathrm{data}}}, \quad
\mathbf{w}_i = [w_i^{(1)},\dots,w_i^{(N_{\mathrm{data}})}]^\top,
\ \mathbf{k}_i(\mathbf{z},Z) = [k_i(\mathbf{z},\mathbf{z}^{(1)}),\dots,k_i(\mathbf{z},\mathbf{z}^{(N_{\mathrm{data}})})]^\top$ and $\mathbf{K}_i$ is the Gram matrix with entries $[\mathbf{K}_i]_{pq} = k_i(\mathbf{z}^{(p)},\mathbf{z}^{(q)}),$ and the (diagonal) covariance is
$\hat{\boldsymbol{\Sigma}}_\mathbf{w}(\mathbf{x},\mathbf{u}) = \operatorname{diag}(\hat{\boldsymbol{\sigma}}_{\mathbf{w},1}^2(\mathbf{x},\mathbf{u}), \ldots, \hat{\boldsymbol{\sigma}}_{\mathbf{w},n}^2(\mathbf{x},\mathbf{u}))$ at any query point $(\mathbf{x},\mathbf{u})$.

\textbf{Deterministic abstraction:}
Rather than propagating GP uncertainty via chance constraints, we wrap GP posteriors in deterministic confidence sets~\cite{hewing2018correspondence}. 
We distinguish three uncertainty types: \textbf{(i)} \emph{state–independent aleatoric} (fixed noise law), \textbf{(ii)} \emph{state–dependent epistemic} (unknown deterministic $g(\mathbf{x},\mathbf{u})$), and \textbf{(iii)} \emph{state–dependent aleatoric} (noise law $\mathcal D_{\mathbf{x},\mathbf{u}}$ varies with $(\mathbf{x},\mathbf{u})$). Most probabilistic invariance assumes \textbf{(i)}, and learning–robust control typically addresses \textbf{(ii)} by set–wrapping GP posteriors. We target \textbf{(iii)}: tubes/invariant sets when the distribution changes with $(x,u)$, using an epistemic shrink–wrap plus a local chance operator to obtain
\begin{equation}
\begin{aligned}
\mathbb{W}_{\text{GP}}(\mathbf{x},\mathbf{u})
&= \bigl\{\mathbf{w}\in\mathbb{R}^n:\,
(\mathbf{w}-\hat{\boldsymbol{\mu}}_\mathbf{w}(\mathbf{x},\mathbf{u}))^\top\\
&\hat{\boldsymbol{\Sigma}}_\mathbf{w}(\mathbf{x},\mathbf{u})^{-1}
(\mathbf{w}-\hat{\boldsymbol{\mu}}_\mathbf{w}(\mathbf{x},\mathbf{u}))
\le \chi^2_{n,1-\alpha}\bigr\},
\end{aligned}
\label{eq:WGP}
\end{equation}
which encloses the true disturbance with probability $1-\alpha$ while enabling standard robust set operations. Here $\chi^2_{n,1-\alpha}$ denotes the $(1-\alpha)$ quantile of the $\chi^2$ distribution with $n$ degrees of freedom, i.e., the smallest $c>0$ such that $\mathbb{P}\{\mathbf{\eta}^\top\mathbf{\eta}\le c\}=1-\alpha$ for $\mathbf{\eta}\sim\mathcal N(0,I_n)$. Equivalently, $(\mathbf{w}-\hat\mu)^\top \hat\Sigma^{-1} (\mathbf{w}-\hat\mu)\le \chi^2_{n,1-\alpha}$ is a highest-density credible ellipsoid under the Gaussian posterior.

\subsection{RPI for State- and Input-Dependent Disturbances}\label{subsec:2C}

The fundamental challenge in computing RPI sets for LTI systems with state- and input-dependent disturbances lies in the circularity of the invariance condition. Consider a control law of the form $\mathbf{u}(k) = \mathbf{K}\mathbf{x}(k) + \mathbf{v}(k)$, where $\mathbf{K} \in \mathbb{R}^{m \times n}$ is a stabilizing feedback gain and $\mathbf{v}(k) \in \mathbb{V} \subseteq \mathbb{R}^m$ is an auxiliary control input. This results in the closed-loop dynamics:
\begin{equation}\label{eq:closed_loop}
    \begin{aligned}
        \mathbf{x}(k+1) &= (\mathbf{A}+\mathbf{BK})\mathbf{x}(k) + \mathbf{B}\mathbf{v}(k) \\ &+ \mathbf{w}(\mathbf{x}(k),\mathbf{K}\mathbf{x}(k)+\mathbf{v}(k)).
    \end{aligned}
\end{equation}
For traditional RPI sets with fixed disturbance sets, the invariance condition has the simple form $\mathbf{A_{\text{cl}}}\Omega \oplus \mathbb{W} \subseteq \Omega$ \cite{kolmanovsky1998theory}, where $\mathbf{A_{\text{cl}}} = \mathbf{A}+\mathbf{BK}$ is the closed-loop system matrix and $\oplus$ is the Minkowski sum, allowing straightforward fixed-point iterations that converge monotonically.

\subsubsection{Fixed-Point Circularity} 

With state- and input-dependent disturbances, the RPI condition becomes:
\begin{equation} \label{eq:rpi_formulation}
\begin{aligned}
        &\forall \mathbf{x} \in \Omega, \exists \mathbf{v} \in \mathbb{V},\\ & \forall \mathbf{w} \in \mathbb{W}(\mathbf{x},\mathbf{Kx}+\mathbf{v}) : \mathbf{A_{\text{cl}}x} + \mathbf{Bv} + \mathbf{w} \in \Omega.
\end{aligned}
\end{equation} 
This introduces a fundamental logical circularity: to verify if $\Omega$ is an RPI set, we need to evaluate $F(\Omega) \subseteq \Omega$, where $F(\Omega) = \bigcup_{\mathbf{x} \in \Omega} \bigcup_{\mathbf{v} \in \mathbb{V}} \{\mathbf{A_{\text{cl}}x} + \mathbf{Bv} + \mathbf{w} : \mathbf{w} \in \mathbb{W}(\mathbf{x},\mathbf{Kx}+\mathbf{v})\}$. The circularity stems from the fact that $F$ itself depends on $\Omega$ through the mapping $F(\Omega) = F(\Omega, \mathbb{W}(\Omega))$, where $\mathbb{W}(\Omega) = \bigcup_{\mathbf{x} \in \Omega, \mathbf{v} \in \mathbb{V}} \mathbb{W}(\mathbf{x},\mathbf{Kx}+\mathbf{v})$.

This self-referential dependency breaks the monotonicity property critical for standard fixed-point iterations. Unlike the state-independent case where $\Omega_{i+1} = \mathbf{A_{\text{cl}}}\Omega_i \oplus \mathbb{W}$ guarantees $\Omega_i \subseteq \Omega_j \implies \Omega_{i+1} \subseteq \Omega_{j+1}$, with state-dependent disturbances we have $\Omega_{i+1} = \mathbf{A_{\text{cl}}}\Omega_i \oplus \mathbb{W}(\Omega_i)$, which may expand or contract non-monotonically between iterations. This undermines convergence guarantees of traditional fixed-point methods and potentially leads to multiple distinct fixed-point solutions for different initial estimates $\Omega_0$.

Past works~\cite{ghaemi2011robust, schaich2015robust, rakovic2003reachability} address this circularity by \emph{lifting} the system into an extended space that explicitly includes both the state and the disturbance (or disturbance parameters). Following \cite{ghaemi2011robust}, we define the control increment $\boldsymbol{\delta}\mathbf{v}(k) = \mathbf{v}(k) - \mathbf{v}(k-1)$ and construct an augmented state vector:
\begin{equation}
\boldsymbol{\xi}(k) = \begin{bmatrix} \mathbf{x}(k) \\ \mathbf{v}(k-1) \\ \mathbf{w}(k-1) \end{bmatrix} \in \mathbb{R}^{2n+m}.
\label{eq:augmented}
\end{equation}

The augmented dynamics in this lifted space become:
\begin{equation}
\begin{aligned}
\small
    \boldsymbol{\xi}(k+1) &= \begin{bmatrix} \mathbf{x}(k+1) \\ \mathbf{v}(k) \\ \mathbf{w}(k) \end{bmatrix}
    \\ &= \begin{bmatrix} \mathbf{A_{\text{cl}}x}(k) + \mathbf{Bv}(k) + \mathbf{w}(k) \\ \mathbf{v}(k-1) + \boldsymbol{\delta}\mathbf{v}(k) \\ \mathbf{w}(k) \end{bmatrix}
\end{aligned}
\end{equation}

This can be written in compact form as:
\begin{equation}
\begin{aligned}
\small
\boldsymbol{\xi}(k+1)
&=
\underbrace{\begin{bmatrix}
\mathbf{A}_{\text{cl}} & \mathbf{B} & \mathbf{0} \\
\mathbf{0}             & \mathbf{I} & \mathbf{0} \\
\mathbf{0}             & \mathbf{0} & \mathbf{0}
\end{bmatrix}}_{\widetilde{\mathbf{A}}}\boldsymbol{\xi}(k)
\;\\&+\;
\underbrace{\begin{bmatrix}
\mathbf{B} \\ \mathbf{I} \\ \mathbf{0}
\end{bmatrix}}_{\widetilde{\mathbf{B}}}\boldsymbol{\delta}\mathbf{v}(k)
\;+\;
\underbrace{\begin{bmatrix}
\mathbf{I} \\ \mathbf{0} \\ \mathbf{I}
\end{bmatrix}}_{\widetilde{\mathbf{D}}}\,\mathbf{w}(k),
\end{aligned}
\label{eq:augmented_dynamics}
\end{equation}
\noindent where $ \boldsymbol{\xi}(k)$ denotes the lifted (augmented) state vector, the augmented system matrix is denoted as $\widetilde{\mathbf{A}} \in \mathbb{R}^{(2n+m) \times (2n+m)}$, $\widetilde{\mathbf{B}} \in \mathbb{R}^{(2n+m) \times m}$ is the augmented control input matrix, and $\widetilde{\mathbf{D}} \in \mathbb{R}^{(2n+m) \times n}$ injects the current disturbance $\mathbf{w}(k)$ into the $x$-update (top block) and stores it in the disturbance memory (bottom block). 

In this augmented system, the state and input dependency of the disturbance is captured by defining the constraint set:
\begin{equation}
\label{eq:graph}
\mathcal{G} \;=\; \bigl\{(\mathbf{x},\mathbf{v},\mathbf{w}) \in \mathbb{R}^{2n+m} :\ \mathbf{w} \in \mathbb{W}(\mathbf{x},\mathbf{Kx}+\mathbf{v})\bigr\},
\end{equation}
\noindent which encodes the state–disturbance coupling explicitly within the state space of the augmented system. It is important to note that while the disturbance $\mathbf{w}$ remains state- and input-dependent, the set $\mathcal{G}$ itself is a fixed subset of the extended state space. This insight allows us to reformulate the problem in a space where standard RPI computation techniques become applicable.

The key advantage of this formulation is that we can now compute an RPI set $Z \subset \mathcal{G}$ for the augmented system,
where the invariance condition becomes:
\begin{equation}
\bigl(\,\widetilde{\mathbf{A}}Z \;\oplus\; \widetilde{\mathbf{B}}\,\Delta\mathbb{V} \;\oplus\; \widetilde{\mathbf{D}}\,\mathbb{W}(\operatorname{Proj}_{\mathbf{x},\mathbf{v}}(Z))\,\bigr)\ \cap\ \mathcal{G}
\ \subseteq\ Z,
\end{equation}
\noindent where $\Delta\mathbb{V} = \{\mathbf{v}_1 - \mathbf{v}_2 : \mathbf{v}_1, \mathbf{v}_2 \in \mathbb{V}\}$ is the set of feasible control increments, $\operatorname{Proj}_{\mathbf{x},\mathbf{v}}(Z)$ denotes the projection of $Z$ onto the $(\mathbf{x},\mathbf{v})$ components, and $\mathbb{W}(\operatorname{Proj}_{\mathbf{x},\mathbf{v}}(Z))$ represents the collection of disturbance sets for all state–input pairs in the projection. While this condition still incorporates the state- and input-dependent disturbance set, the formulation within the augmented state space allows us to apply fixed-point methods in a well-defined manner.

To implement this approach, a two-stage approximation process is employed: first, we convert GP predictions to confidence ellipsoids, as described in Section \ref{subsec:2B}; then, we approximate these ellipsoids with polytopes to enable efficient set operations. Our method transitions systematically from worst-case to data-driven disturbance bounds. Initially, we employ conservative bounds to ensure safety in poorly explored regions. As the GP model refines with additional data, posterior variance reduction enables dynamic shrinking of these bounds, maintaining formal safety guarantees while progressively reducing conservatism~\cite{koller2018learning}.

\section{Learning-Based Robust Control Synthesis} \label{sec:Synthesis}

This section presents our integrated methodology, building on the GP-based disturbance modeling framework established in Section \ref{sec:RPI_compute}, we develop a comprehensive control synthesis approach that addresses the circular dependency in RPI computations while leveraging learned uncertainty bounds for reduced conservatism.

\subsection{From GP Ellipsoids to Polytopic Control Constraints}

The GP framework from Section \ref{sec:RPI_compute} yields confidence ellipsoids $\mathbb{W}_{\text{GP}}$ in~\eqref{eq:WGP}
that provide probabilistic guarantees but are incompatible with standard MPC optimization frameworks. Robust control synthesis commonly requires deterministic polytopic bounds that enable linear constraint formulations and efficient set operations (Minkowski sums, intersections, projections) essential for RPI computation \cite{7431977}.

We address this through polytopic outer approximation: for each ellipsoid $\mathbb{W}_{\text{GP}}(\mathbf{x},\mathbf{u})$, we construct a polytope $\mathbb{W}_{\text{poly}}(\mathbf{x},\mathbf{u}) = \{\mathbf{w} \in \mathbb{R}^n : \mathbf{H_w w} \leq \mathbf{h_w}\}$ such that $\mathbb{W}_{\text{GP}}(\mathbf{x},\mathbf{u}) \subseteq \mathbb{W}_{\text{poly}}(\mathbf{x},\mathbf{u})$, where $\mathbf{H_w} \in \mathbb{R}^{n_f \times n}$ and $\mathbf{h_w} \in \mathbb{R}^{n_f}$ define the polytopic constraints with $n_f$ facets. This conversion preserves the original probabilistic guarantees while enabling computational tractability: if $\mathbb{P}[\mathbf{w}(\mathbf{x},\mathbf{u}) \in \mathbb{W}_{\text{GP}}(\mathbf{x},\mathbf{u})] = 1-\alpha$, then $\mathbb{P}[\mathbf{w}(\mathbf{x},\mathbf{u}) \in \mathbb{W}_{\text{poly}}(\mathbf{x},\mathbf{u})] \geq 1-\alpha$.

Critically, the polytopic approximation preserves the adaptive sizing properties of GP uncertainty quantification. In data-dense regions where $\hat{\boldsymbol{\Sigma}}_\mathbf{w}(\mathbf{x},\mathbf{u})$ is small, the resulting polytopes $\mathbb{W}_{\text{poly}}(\mathbf{x},\mathbf{u})$ are correspondingly tight, reducing conservatism. In unexplored regions where $\hat{\boldsymbol{\Sigma}}_\mathbf{w}(\mathbf{x},\mathbf{u})$ approaches prior values, larger polytopes maintain robust safety margins.

\subsection{Lift–and–Project Framework for RPI Computation}
\label{ssec:lift_project}

State– and input–dependent disturbances make the RPI test circular: the set to be verified depends on itself. We break this by a lift–and–project formulation that augments the state with disturbance variables and encodes the coupling as a fixed graph constraint $\mathcal G$ in the extended space; the RPI search then reduces to a standard fixed-point computation under this static constraint. \textbf{Two nested time scales:}
We separate \textbf{(i)} \emph{learning epochs} $q$, where the GP posterior and its polytope $\widehat{\mathbb W}^{(q)}$ are frozen, from \textbf{(ii)} an \emph{inner fixed-point iteration} $k=0,1,\dots$ that computes the RPI set for that frozen description. The index $k$ is not physical time. When new data arrive, GP variance contracts so $\widehat{\mathbb W}^{(q+1)}\subseteq\widehat{\mathbb W}^{(q)}$; we warm-start a fresh fixed-point run from the previous solution. This separation preserves rigor while accommodating evolving uncertainty. The lift–and–project framework is summarized in Algorithm~\ref{alg:stmpc_epoch}.

\subsubsection{Invariance in the Lifted Space}
We work with the closed-loop dynamics~\eqref{eq:closed_loop} and the augmented state $\boldsymbol{\xi}=\bigl[\mathbf{x}^\top,\;\mathbf{v}^\top,\;\mathbf{w}^\top\bigr]^\top$ of~\eqref{eq:augmented}. The algebraic coupling $\mathbf{w}\in\mathbb{W}(\mathbf{x},\mathbf{u})$ is encoded by the fixed graph set
\(
\mathcal{G}\!=\!\bigl\{(\mathbf{x},\mathbf{v},\mathbf{w}):
           \mathbf{w}\!\in\!\mathbb{W}(\mathbf{x},\mathbf{K}\mathbf{x}+\mathbf{v})\bigr\}.
\)

\subsection*{Assumptions \& Regularity}
\begin{enumerate}
\item (Asm. 2) \textbf{Domain/constraints:} $\mathbb X\subset\mathbb R^n$, $\mathbb U,\mathbb V\subset\mathbb R^m$ are non-empty, compact, convex; $K$ renders $A{+}BK$ Schur.
\item (Asm. 3) \textbf{Disturbance map:} $(x,u)\mapsto \widehat{\mathbb W}(x,u)$ is non-empty, compact-valued, upper hemicontinuous, with closed graph; its polyhedralization preserves closed graph.
\item (Asm. 4) \textbf{Lifted graph:} $\mathcal G=\{(x,v,w): w\in \widehat{\mathbb W}(x,Kx{+}v)\}$ is closed, and all lifted images under $\tilde A,\tilde B,\tilde D$ remain bounded in $\mathcal G$.
\item  (Asm. 5) \textbf{Auxiliary input set:} $\mathbb{V}\subseteq\mathbb{R}^m$ is compact and convex, and we denote $\Delta\mathbb{V}=\{\mathbf{v}_1-\mathbf{v}_2: \mathbf{v}_{1,2}\in\mathbb{V}\}$.
\item \textbf{Supports:} support directions $\mathcal S$ are fixed, finite, and bounded.

\end{enumerate}

\paragraph{Forward outside-in operator.}
For any $Z\subseteq\mathcal{G}$ define
\begin{equation}
\mathcal{F}(Z)\;:=\;\bigl(\widetilde{\mathbf{A}}Z
          \,\oplus\,\widetilde{\mathbf{B}}\,\Delta\mathbb{V}
          \,\oplus\,\widetilde{\mathbf{D}}\,W(Z)\bigr)\;\cap\;\mathcal{G},
\label{eq:F_operator_new}
\end{equation}
with
\(
W(Z)=\!\bigcup\limits_{(\mathbf{x},\mathbf{v})\in\operatorname{Proj}_{\mathbf{x},\mathbf{v}}(Z)}
      \mathbb{W}(\mathbf{x},\mathbf{K}\mathbf{x}+\mathbf{v}).
\)
A set $Z^\star$ is RPI for the lifted system iff $\mathcal{F}(Z^\star)\subseteq Z^\star$. Because we enforce $\mathcal{F}(Z_0)\subseteq Z_0$ at the start of every epoch, the sequence $Z_{k+1}=\mathcal{F}(Z_k)$ \emph{shrinks}, i.e.\ $Z_{k+1}\subseteq Z_k$, hence ``outside-in''.

\subsection*{Monotonicity}
\begin{lemma}[Monotonicity property]
\label{lem:monotonicity_new}
If $Z_1\subseteq Z_2\subseteq\mathcal{G}$, then
$\mathcal{F}(Z_1)\subseteq\mathcal{F}(Z_2)$.
\end{lemma}
\begin{proof}
Assume $Z_1 \subseteq Z_2$. Since $\widetilde{\mathbf{A}}$ is linear, 
$\widetilde{\mathbf{A}}Z_1\subseteq\widetilde{\mathbf{A}}Z_2$ follows immediately.
The projection satisfies $\operatorname{Proj}_{\mathbf{x},\mathbf{v}}(Z_1) \subseteq \operatorname{Proj}_{\mathbf{x},\mathbf{v}}(Z_2)$; 
hence every pair $(\mathbf{x},\mathbf{v})$ that contributes to $W(Z_1)$ also contributes to $W(Z_2)$, 
giving $W(Z_1)\subseteq W(Z_2)$. Because $\widetilde{\mathbf{D}}$ is linear, 
$\widetilde{\mathbf{D}}W(Z_1)\subseteq\widetilde{\mathbf{D}}W(Z_2)$ follows.
The Minkowski sum preserves inclusions:
\begin{equation*} \begin{aligned}
  \widetilde{\mathbf{A}}Z_1 \,\oplus\, \widetilde{\mathbf{B}}\Delta\mathbb{V} \,\oplus\, \widetilde{\mathbf{D}}W(Z_1) 
  \subseteq \widetilde{\mathbf{A}}Z_2 \,\oplus\, \widetilde{\mathbf{B}}\Delta\mathbb{V} \,\oplus\, \widetilde{\mathbf{D}}W(Z_2).
\end{aligned} \end{equation*} 
Intersecting both sides with $\mathcal{G}$ preserves the inclusion, yielding $\mathcal{F}(Z_1)\subseteq\mathcal{F}(Z_2)$.
\end{proof}

\subsection*{Outside-in convergence (single epoch)}
\begin{lemma}[Cantor–Bolzano fixed point]
\label{lem:convergence_new}
Choose $Z_0\subseteq\mathcal{G}$ non-empty, compact and such that
$\mathcal{F}(Z_0)\subseteq Z_0$.
Then the decreasing chain
$Z_{k+1}=\mathcal{F}(Z_k)$ satisfies
\begin{enumerate}\itemsep2pt
\item $Z_{k+1}\subseteq Z_k$ for all $k$ (outside-in),
\item $Z_\infty:=\bigcap_{k=0}^\infty Z_k$ is non-empty and compact, and
\item $Z_\infty=\mathcal{F}(Z_\infty)$.
\end{enumerate}
\end{lemma}
\begin{proof}
\textbf{Item 1:} We have $Z_1 = \mathcal{F}(Z_0) \subseteq Z_0$ by assumption. 
By Lemma~\ref{lem:monotonicity_new}, $Z_2 = \mathcal{F}(Z_1) \subseteq \mathcal{F}(Z_0) = Z_1$. 
By induction, $Z_{k+1} = \mathcal{F}(Z_k) \subseteq Z_k$ for all $k \geq 0$. \textbf{Item 2:} The sequence $Z_0 \supseteq Z_1 \supseteq Z_2 \supseteq \ldots$ 
is a decreasing chain of non-empty compact sets. By Cantor's intersection theorem, 
$Z_\infty = \bigcap_{k=0}^\infty Z_k$ is non-empty and compact. \textbf{Item 3:} To show $Z_\infty = \mathcal{F}(Z_\infty)$, we prove both inclusions.
For $\mathcal{F}(Z_\infty) \subseteq Z_\infty$: Since $Z_\infty \subseteq Z_k$ for all $k$, Lemma~\ref{lem:monotonicity_new} gives $\mathcal{F}(Z_\infty) \subseteq \mathcal{F}(Z_k) = Z_{k+1}$ for all $k$. Hence $\mathcal{F}(Z_\infty) \subseteq \bigcap_{k=0}^\infty Z_{k+1} = Z_\infty$. For $Z_\infty \subseteq \mathcal{F}(Z_\infty)$: Take any $\boldsymbol{\xi} \in Z_\infty$. Then $\boldsymbol{\xi} \in Z_{k+1} = \mathcal{F}(Z_k)$ for all $k \geq 0$. For each $k$, there exists $\boldsymbol{\zeta}_k \in Z_k$ such that $\boldsymbol{\xi} \in \mathcal{F}(\{\boldsymbol{\zeta}_k\})$. Since $\{\boldsymbol{\zeta}_k\}_{k=0}^\infty \subseteq Z_0$ (compact), by Bolzano–Weierstrass, a subsequence $\{\boldsymbol{\zeta}_{k_j}\}$ converges to some $\boldsymbol{\zeta}^* \in Z_\infty$. By the closed-graph property of $\mathcal{F}$ (guaranteed by upper hemicontinuity of $\mathbb{W}$ and compactness), as $\boldsymbol{\zeta}_{k_j} \to \boldsymbol{\zeta}^*$ with $\boldsymbol{\xi} \in \mathcal{F}(\{\boldsymbol{\zeta}_{k_j}\})$, we have $\boldsymbol{\xi} \in \mathcal{F}(\{\boldsymbol{\zeta}^*\}) \subseteq \mathcal{F}(Z_\infty)$.
\end{proof}

\subsection*{Projection back to the plant coordinates}
\begin{lemma}[RPI via measurable selector]
\label{lem:rpi_selector}
Let $Z^\star\subseteq \mathcal G$ be a lifted fixed point for a frozen epoch and define 
$\mathbb Z^\star := \mathrm{Proj}_x(Z^\star)$. Suppose $(x,u)\mapsto \widehat{\mathbb W}(x,u)$ has closed graph and compact values. Then there exists a Borel-measurable selector $\kappa:\mathbb Z^\star\to\mathbb V$ such that
$(\!A{+}BK\!)x + B\,\kappa(x) + w \in \mathbb Z^\star \quad \forall x\in\mathbb Z^\star,\ \forall w\in \widehat{\mathbb W}(x,Kx{+}\kappa(x)).$ Hence $\mathbb Z^\star$ is RPI for the plant.
\end{lemma}

\begin{proof}
Let $T(\mathbf{x}):=\{\,\mathbf{v}\in\mathbb V:\ \exists\,\mathbf{w}\ \text{s.t.}\ (\mathbf{x},\mathbf{v},\mathbf{w})\in Z^\star\,\}$; then $T$ has non-empty compact values and closed (hence Borel) graph $\operatorname{Graph}(T)=\operatorname{Proj}_{x,v}(Z^\star)$. By the Kuratowski--Ryll-Nardzewski measurable selection theorem, there exists a Borel selector $\boldsymbol{\kappa}(\mathbf{x})\in T(\mathbf{x})$. For any $\mathbf{x}$ and any $\mathbf{w}\in\widehat{\mathbb W}(\mathbf{x},\mathbf{K}\mathbf{x}+\boldsymbol{\kappa}(\mathbf{x}))$, taking $\boldsymbol{\delta v}=\mathbf{0}$ and using $Z^\star=\mathcal F(Z^\star)$ yields $\big((\mathbf{A}{+}\mathbf{B}\mathbf{K})\mathbf{x}+\mathbf{B}\boldsymbol{\kappa}(\mathbf{x})+\mathbf{w},\ \boldsymbol{\kappa}(\mathbf{x}),\ \mathbf{w}\big)\in Z^\star$, hence $(\mathbf{A}{+}\mathbf{B}\mathbf{K})\mathbf{x}+\mathbf{B}\boldsymbol{\kappa}(\mathbf{x})+\mathbf{w}\in\mathbb Z^\star$.
\end{proof}

\subsection*{Maintaining Uniform Safety through $\varepsilon$-nets}
\begin{lemma}[Uniform safety of Anchors]\label{lem:uniform_safety_compact}
Let $\mathcal R_{x,v}\subset\mathbb X\times\mathbb V$ be compact and let $\mathcal A_{x,v}$ be an $\varepsilon$-net (finite grid covering) of $\mathcal R_{x,v}$.
For $(\mathbf x,\mathbf v)\in\mathcal R_{x,v}$ set $\mathbf u=\mathbf K\mathbf x+\mathbf v$ and $\mathbf z=(\mathbf x,\mathbf u)$; for each anchor $(\mathbf x_a,\mathbf v_a)\in\mathcal A_{x,v}$ set $\mathbf u_a=\mathbf K\mathbf x_a+\mathbf v_a$ and $\mathbf z_a=(\mathbf x_a,\mathbf u_a)$.
Fix $\alpha_{\mathrm{anc}}\in(0,1)$ and define $c_{n,\alpha}:=\sqrt{\chi^2_{n,\,1-\alpha}}$.
Let $\mathcal S\subset\mathbb R^n$ be the fixed finite set of support directions (unit vectors) used to define the disturbance polytopes. For each $\mathbf{s}\in\mathcal S$ define
$
h(\mathbf z;\mathbf{s})
= \mathbf{s}^\top\hat{\boldsymbol\mu}_\mathbf w(\mathbf z)
 \;+\;
 c_{n,\alpha_{\mathrm{anc}}}\ \sigma_\mathbf{s}(\mathbf z),\
\sigma_\mathbf{s}(\mathbf z):=\sqrt{\mathbf{s}^\top\hat{\boldsymbol\Sigma}_\mathbf w(\mathbf z)\,\mathbf{s}}.
$
Assume $\hat{\boldsymbol\mu}_\mathbf w$ is $L_\mu$–Lipschitz and, for each fixed $\mathbf{s}\in\mathcal S$, $\sigma_\mathbf{s}$ is $L_\sigma$–Lipschitz on
$\mathcal R_{x,u}=\{(\mathbf x,\mathbf K\mathbf x+\mathbf v):(\mathbf x,\mathbf v)\in\mathcal R_{x,v}\}$.
Define the anchor envelope with Lipschitz inflation
\begin{equation}
\begin{aligned}
\overline h(\mathbf{s})\ &:=\
\underbrace{\max_{(\mathbf x_a,\mathbf v_a)\in\mathcal A_{x,v}}
\Bigl[\mathbf{s}^\top\hat{\boldsymbol\mu}_\mathbf w(\mathbf{z}_a)
      +c_{n,\alpha_{\mathrm{anc}}}\ \sigma_\mathbf{s}(\mathbf{z}_a)\Bigr]}_{\text{anchor supports}}
\\[0.25em]
&+\;
\underbrace{\bigl(\|\mathbf{s}\|L_\mu+c_{n,\alpha_{\mathrm{anc}}}L_\sigma\bigr)\,\varepsilon}_{\text{Lipschitz inflation}}.
\end{aligned}
\end{equation}
Then, for all $(\mathbf x,\mathbf v)\in\mathcal R_{x,v}$ (hence $\mathbf{z}\in\mathcal R_{x,u}$) and all $\mathbf s\in\mathcal S$,
$
h(\mathbf{z};\mathbf{s}) \;\le\; \overline h(\mathbf{s}).
$
Moreover, if each anchor ellipsoid is $(1-\alpha_{\mathrm{anc}})$-credible under the GP posterior, then by the union bound the probability that
\emph{some} anchor ellipsoid fails to contain the true disturbance is at most
$
\alpha_{\mathrm{uniform}} \;:=\; |\mathcal A_{x,v}|\ \alpha_{\mathrm{anc}}.
$
Consequently, using
$
\varepsilon_{\mathrm{cov}}(\mathbf{s})
=
\bigl(\|\mathbf{s}\|L_\mu+c_{n,\alpha_{\mathrm{anc}}}L_\sigma(\mathbf{s})\bigr)\,\varepsilon
$
in Algorithm~\ref{alg:stmpc_epoch} yields a uniform outer wrapper whose probability of excluding the true disturbance at some anchor (in some $\mathbf s\in\mathcal S$) is at most $\alpha_{\mathrm{uniform}}$.
\end{lemma}

\begin{proof}[Proof sketch]
Claim follows directly from the per–anchor $(1-\alpha_{\mathrm{anc}})$ credibility of the GP ellipsoids, the union bound over the finite net $\mathcal A_{x,v}$, and the Lipschitz bounds on $\hat{\boldsymbol\mu}_\mathbf w$ and $\sigma_\mathbf{s}$; the detailed derivation is therefore omitted for brevity.
\end{proof}

\subsection{Main Result for a Single Learning Epoch}

\begin{theorem}[RPI existence, projection, and per-epoch uniform safety]\label{thm:RPI_epoch}
Fix an epoch $q$ with a frozen wrapper $\widehat{\mathbb W}^{(q)}(\cdot)$ constructed from GP posteriors as in Section~\ref{sec:RPI_compute}, polyhedralized on a fixed finite set of supports, and uniformized as in Lemma~\ref{lem:uniform_safety_compact} with risk budget $\alpha_{\mathrm{epoch}}\in(0,1)$ over a compact design domain $\mathcal R_{x,v}$.
Let $\mathcal{G}=\{(\mathbf{x},\mathbf{v},\mathbf{w}):\ \mathbf{w}\in \widehat{\mathbb W}^{(q)}(\mathbf{x},\mathbf{K}\mathbf{x}+\mathbf{v})\}$ and define the isotone outside–in operator in ($\ref{eq:F_operator_new}$)
\begin{equation*} \begin{aligned} 
W(Z)&=\!\!\!\bigcup_{(\mathbf{x},\mathbf{v})\in\operatorname{Proj}_{\mathbf{x},\mathbf{v}}(Z)}\!\!\!\!\!\widehat{\mathbb W}^{(q)}(\mathbf{x},\mathbf{K}\mathbf{x}+\mathbf{v}).
\end{aligned} \end{equation*} 
Assume $\mathbf{A}{+}\mathbf{B}\mathbf{K}$ is Schur, $\mathbb X,\mathbb U,\mathbb V$ are compact convex, and $(\mathbf{x},\mathbf{u})\mapsto \widehat{\mathbb W}^{(q)}(\mathbf{x},\mathbf{u})$ has compact values and closed graph on $\mathcal R_{x,u}$. Pick a non-empty compact $Z_0\subseteq \mathcal G$ with $\mathcal F(Z_0)\subseteq Z_0$. Then: (\emph{Existence and convergence}) The sequence $Z_{k+1}=\mathcal F(Z_k)$ is decreasing, $Z_{k+1}\subseteq Z_k$, and converges (in the Painlev\'e--Kuratowski sense) to a non-empty compact fixed point $Z^\star=\mathcal F(Z^\star)\subseteq\mathcal G$. (\emph{Plant-level invariance}) There exists a Borel-measurable selector $\boldsymbol{\kappa}:\mathbb Z^\star\to\mathbb V$ such that, with $\mathbb Z^\star:=\operatorname{Proj}_{\mathbf{x}}(Z^\star)$,
$
(\mathbf{A}{+}\mathbf{B}\mathbf{K})\mathbf{x}+\mathbf{B}\,\boldsymbol{\kappa}(\mathbf{x})+\mathbf{w}\ \in\ \mathbb Z^\star \\ 
\forall \mathbf{x}\in\mathbb Z^\star,\ \forall \mathbf{w}\in \widehat{\mathbb W}^{(q)}\bigl(\mathbf{x},\mathbf{K}\mathbf{x}+\boldsymbol{\kappa}(\mathbf{x})\bigr).
$
Hence $\mathbb Z^\star$ is RPI for the plant with respect to the epoch outer wrapper $\widehat{\mathbb W}^{(q)}(\cdot)$. (Per-epoch uniform safety) With probability at least $1-\alpha_{\mathrm{epoch}}$ (over the GP posterior within epoch $q$), the true disturbance satisfies
$\mathbf{w}(\mathbf{x},\mathbf{u})\in \widehat{\mathbb W}^{(q)}(\mathbf{x},\mathbf{u})$ \emph{simultaneously for all} $(\mathbf{x},\mathbf{u})\in\mathcal R_{x,u}$; consequently, with the same probability, $\mathbb Z^\star$ is RPI for the true plant disturbances throughout the epoch. (Epoch nesting) If the next-epoch wrapper tightens, $\widehat{\mathbb W}^{(q+1)}\subseteq \widehat{\mathbb W}^{(q)}$, then the corresponding fixed points satisfy $Z^{\star,(q+1)}\subseteq Z^{\star,(q)}$ and $\operatorname{Proj}_{\mathbf{x}}(Z^{\star,(q+1)})\subseteq \operatorname{Proj}_{\mathbf{x}}(Z^{\star,(q)})$. 

\end{theorem}

\begin{proof}
\emph{\textbf{Convergence.}} By Lemma~\ref{lem:monotonicity_new}, $\mathcal F$ is isotone. The initialization $\mathcal F(Z_0)\subseteq Z_0$ makes the sequence $Z_{k+1}=\mathcal F(Z_k)$ decreasing. The sets are nonempty and uniformly bounded (compactness of $\mathbb X,\mathbb V$ and of the values of $\widehat{\mathbb W}^{(q)}$, plus Schur stability of $\mathbf A{+}\mathbf B \mathbf K$ ensure bounded images), hence by Cantor’s theorem the intersection $Z^\star:=\bigcap_k Z_k$ is nonempty and compact. Closedness of the graph of $\mathcal F$ (induced by linearity, Minkowski sums, and the closed graph of $\widehat{\mathbb W}^{(q)}$) yields $Z^\star=\mathcal F(Z^\star)$ as in Lemma~\ref{lem:convergence_new}. \emph{\textbf{Plant-level invariance.}} By Lemma~\ref{lem:rpi_selector}, closed-graph and compact-valuedness of $(\mathbf{x},\mathbf{u})\mapsto \widehat{\mathbb W}^{(q)}(\mathbf{x},\mathbf{u})$ imply the existence of a Borel selector $\kappa(\cdot)$ so that the projection $\mathbb Z^\star$ is RPI for the lifted disturbance wrapper, establishing the claim. \emph{\textbf{Per-epoch uniform safety.}} By Lemma~\ref{lem:uniform_safety_compact} and the choice $\alpha_{\mathrm{anc}}=\alpha_{\mathrm{epoch}}/|\mathcal A_{x,v}|$, with probability at least $1-\alpha_{\mathrm{epoch}}$ we have the \emph{simultaneous} inclusion $\mathbf{w}(\mathbf{x},\mathbf{u})\in \widehat{\mathbb W}^{(q)}(\mathbf{x},\mathbf{u})$ for all $(\mathbf{x},\mathbf{u})\in\mathcal R_{x,u}$ during epoch $q$. The fixed-point $Z^\star$ was computed against this (outer) wrapper; by monotonicity, invariance for the outer wrapper implies invariance for the true (smaller) disturbances, uniformly over $\mathcal R_{x,v}$ within the epoch. \emph{\textbf{Epoch nesting.}} If $\widehat{\mathbb W}^{(q+1)}\subseteq \widehat{\mathbb W}^{(q)}$, then for any $Z$ we have $\mathcal F_{q+1}(Z)\subseteq \mathcal F_q(Z)$. Applying Lemma~\ref{lem:monotonicity_new} to the two operators yields $Z^{\star,(q+1)}\subseteq Z^{\star,(q)}$ and the corresponding inclusion of state projections.
\end{proof}

\section{Simulation and Results} \label{sec:simulations}
We evaluate on a 2D double integrator with $x=[p_x,p_y,v_x,v_y]^\top$, $u=[a_x,a_y]^\top$, and additive, state- and input–dependent disturbances on acceleration,
\[
\begin{bmatrix} w_x \\ w_y \end{bmatrix}
= -\frac{\beta_1}{m}\|\mathbf v\|\!\begin{bmatrix} v_x \\ v_y \end{bmatrix}
  -\frac{\beta_2}{m}\!\begin{bmatrix} u_x \\ u_y \end{bmatrix}
  + \mathbf\vartheta ,
\]
where $\|\mathbf{v}\|$ is the magnitude of the velocity vector in meters per second (m/s), $m$ is the mass of the system in kg, $\beta_1$ is the aerodynamic drag coefficient in kg/m, $\beta_2$ is the actuator efficiency coefficient in kg, and $\mathbf{\vartheta } = [\vartheta _x, \vartheta _y]^T$ is the process noise vector in m/s$^2$.

\begin{figure}[!t]
\centering
\subfloat[$w_x$ over $(v_x,u_x)$]{%
    \includegraphics[width=0.24\textwidth]{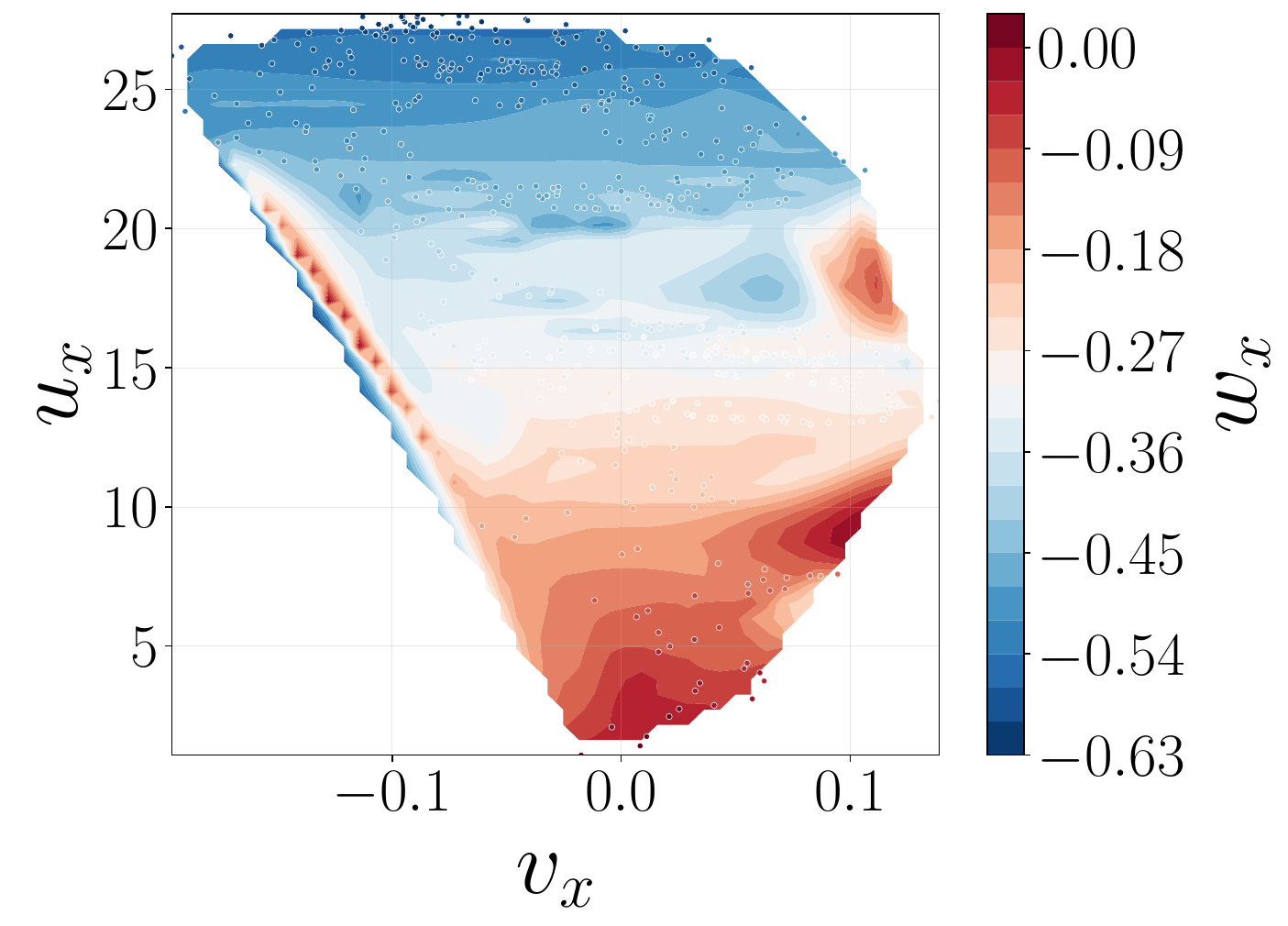}%
    \label{fig:gp_a}%
}\hfill
\subfloat[$w_y$ over $(u_x,u_y)$]{%
    \includegraphics[width=0.24\textwidth]{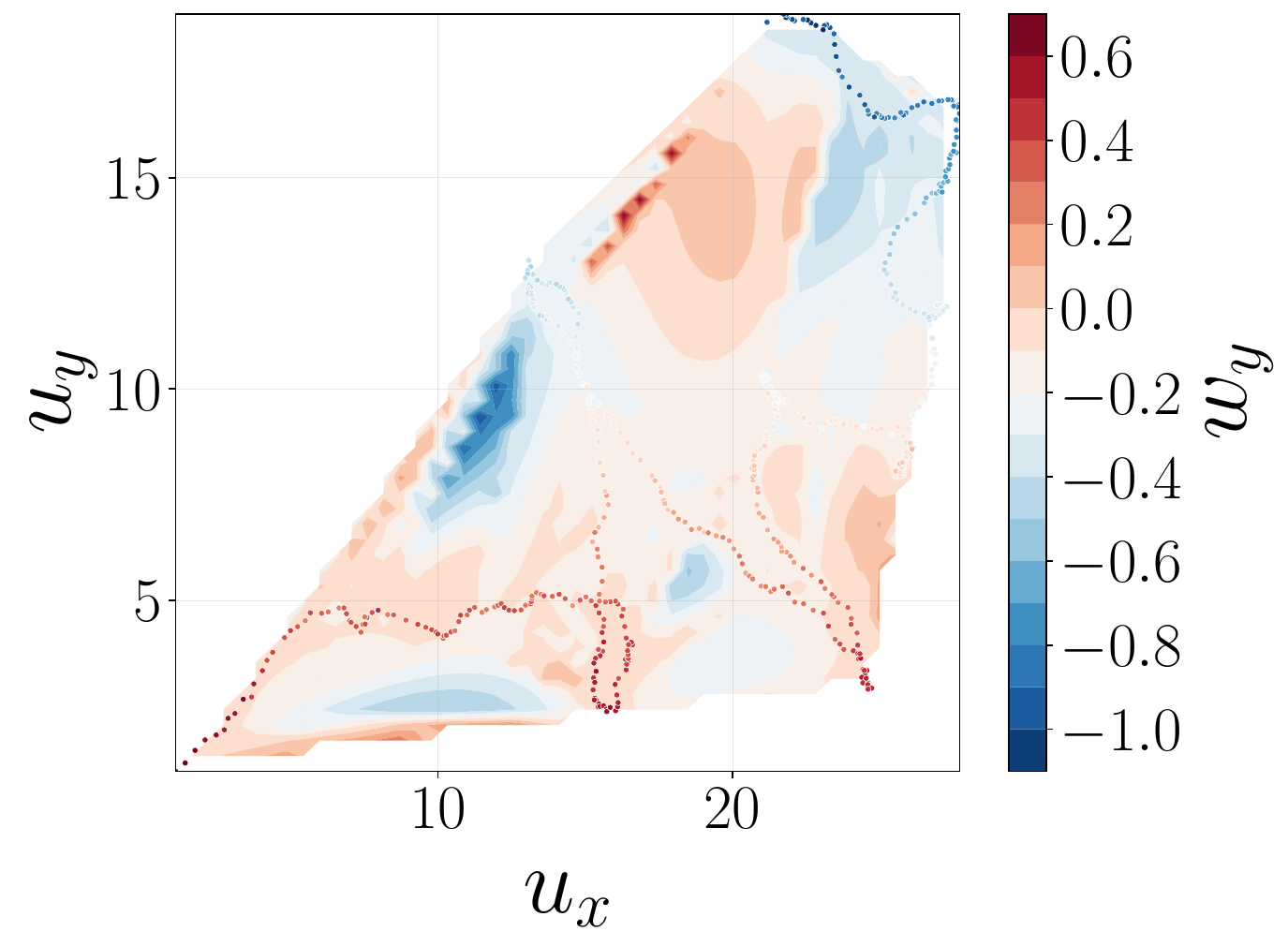}%
    \label{fig:gp_b}%
}\hfill
\subfloat[$(v_x,u_x,w_x)$ slice]{%
    \includegraphics[width=0.24\textwidth]{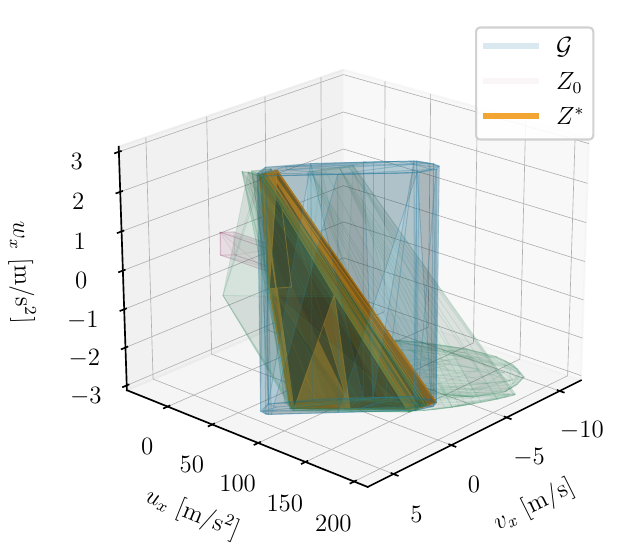}%
    \label{fig:rpi_c}%
}\hfill
\subfloat[$(p_x,u_x,w_x)$ slice]{%
    \includegraphics[width=0.24\textwidth]{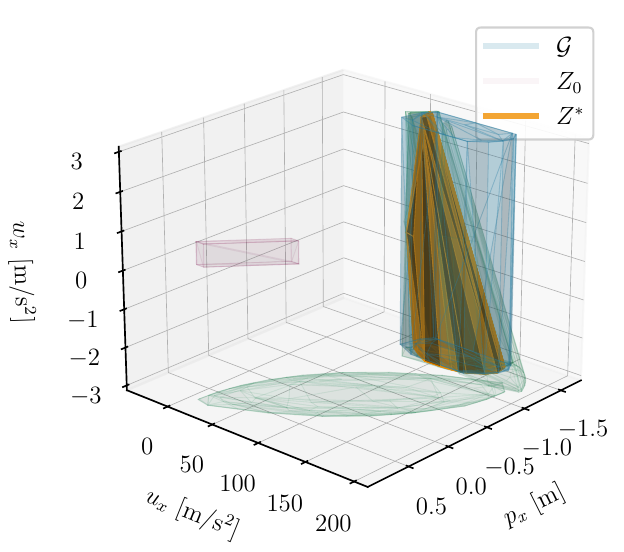}%
    \label{fig:rpi_d}%
}
\caption{ GP–learned disturbance structure and lifted–space RPI. 
\textbf{(a,b)} GP posteriors expose state- and input–dependent effects used to size local disturbance sets. \textbf{(c,d)} The lift–and–project iteration converges to a compact invariant set (yellow) contained in the graph constraint $\mathcal{G}$ (blue). The final projected RPI facet count is ${\operatorname{Proj}_{\mathbf{x}}(Z^{\star,(q)})}_{n_f} = 1466$, convergence metric: Hausdorff distance}
\label{fig:combined_gp_rpi}
\end{figure}

\textbf{Disturbance learning.}
Figures~\ref{fig:combined_gp_rpi}(a,b) illustrate representative slices: velocity-dependent drag in $(v_x,u_x)$ and input coupling in $(u_x,u_y)$. 
These learned structures tighten the local disturbance polytopes where data are informative, reducing conservatism while preserving hard safety via the fixed-point tube. In our bounded domain these coincide; we report $Z^\star$. Figures~\ref{fig:combined_gp_rpi}(c,d) show the contraction to $Z^\star$ (yellow) entirely within the graph constraint (blue). 
Projecting back gives $\operatorname{Proj}_x(Z^\star)$, which is RPI for the plant and provides tube cross-sections for MPC. Our verifiably safe learning approach demonstrates 22.9$\times$ improved accuracy over traditional fixed bounds, reducing overall conservatism by 55.4\% compared to worst-case methods while maintaining safety guarantees.

\begin{algorithm}[htbp!]
\caption{STMPC epoch loop}
\label{alg:stmpc_epoch}
\begin{algorithmic}[1]
\Require dataset $\mathcal D_q$, nominal $(\mathbf{A},\mathbf{B})$, sets $(\mathbb X,\mathbb U,\mathbb V)$, gain $\mathbf{K}$, warm start $Z^{\star,(q-1)}$ (optional)
\State \textbf{Train GP:} fit independent GPs for $\mathbf{w}(\cdot,\cdot)$ on $\mathcal D_q$ to get $\hat{\boldsymbol\mu}_\mathbf{w}(\mathbf{x},\mathbf{u})$, $\hat{\boldsymbol\Sigma}_\mathbf{w}(\mathbf{x},\mathbf{u})$.
\State \textbf{Ellipsoid $\to$ Polytope:} form $\mathbb E(\mathbf{x},\mathbf{u})$ at level $1-\alpha$; outer-approximate by $\widehat{\mathbb W}(\mathbf{x},\mathbf{u})$ (facet-limited).
\State \textbf{Lifted graph:} set $\mathcal G=\{(\mathbf{x},\mathbf{v},\mathbf{w}):\, \mathbf{w}\in\widehat{\mathbb W}(\mathbf{x},\mathbf{K}\mathbf{x}+\mathbf{v})\}$; initialize $Z_0\subseteq\mathcal G$ (warm-start $Z^{\star,(q-1)}$).
\State \textbf{Outside-in:} iterate $Z_{k+1}=\big(\widetilde{\mathbf{A}} Z_k\oplus \widetilde{\mathbf{B}}\,\Delta\mathbb V\oplus \widetilde{\mathbf{D}}\,W(Z_k)\big)\cap\mathcal G$ until gap $<\varepsilon$.
\State \textbf{Projection:} set $Z^{\star,(q)}=\mathrm{fix}(Z_k)$ and $\mathbb Z^{\star,(q)}=\operatorname{Proj}_{\mathbf{x}}(Z^{\star,(q)})$; choose measurable selector $\boldsymbol{\kappa}(\mathbf{x})\in\mathbb V$.
\State \textbf{MPC:} tighten with $\mathbb Z^{\star,(q)}$ and apply $\mathbf{u}=\mathbf{K}\mathbf{x}+\boldsymbol{\kappa}(\mathbf{x})$.
\State \textbf{Repeat:} acquire new data, update $\mathcal D_{q+1}$; set $q\!\leftarrow\! q{+}1$ and repeat from Step~1 (warm-start $Z_0\!\gets\!Z^{\star,(q-1)}$).
\end{algorithmic}
\end{algorithm}

\section{Conclusion}
\label{sec:conclusion}

We presented a learning-based shrinking disturbance invariant scheme that couples with tube MPCs, which learns state- and input-dependent disturbances with GPs and certifies safety through a lifted, order–preserving outside–in fixed–point. Two-time-scale operation (frozen “epochs” for learning versus inner fixed-point iterations) resolves circularity and yields \emph{epoch-to-epoch nesting} of tubes as uncertainty contracts. A double-integrator study illustrates how data tighten local disturbance polytopes and shrink tube cross-sections without relaxing hard constraints. We also detailed a uniform-safety construction over anchor grids. Future work should address the practical computation of the guaranteed measurable selector policy and the robust estimation of Lipschitz constants for uniform safety bounds. A key extension is to handle temporally correlated (colored) disturbances, requiring an augmented state to model disturbance dynamics.

\bibliographystyle{ieeetr}
\bibliography{ref}

\end{document}